# Technical report: Functional Constraint Extraction From Register Transfer Level for ATPG
Christelle Hobeika, Claude Thibeault and Jean-François Boland

We proposed in [1] an automatic functional constraint extractor that can be applied on the RT level. These functional constraints are used to generate pseudo functional test patterns with ATPG tools. The patterns are then used to improve the verification process.

This technical report complements the work proposed in [1] as it contains the implementation details of the proposed methodology and shows the detailed intermediate and final results of the application of this methodology on a concrete example.

## I. HIGH LEVEL STATE DEFINITION

In our work, we consider two types of states: the High abstraction Level State (HLS) and the Low abstraction Level or RTL state (simply called *state* in the rest of the paper) related to the state signals. We define a high level state as the set of RTL state signal assignments associated with particular conditions within a process. Thus, if for a given condition, the value of a counter (state signal) is incremented, then the HLS contains all the possible values that this counter can have, but each incremented value of this same counter is considered as an RTL state.

We define HLS as the set of state signal ranges of values that can appear simultaneously in the design under the same conditions. It corresponds to the values of internal state signals, inputs and outputs that can occur simultaneously with:

$$\text{HLS}_{id\_hlstate} = \bigcup_{j=1}^{N} \text{range of values } x_j \text{, and } x_j \text{ is a state signal}$$

$x_j \in \{In_{id\_module}, IS_{id\_module}, Out_{id\_module}\}$ where:

- $In_{id\_module}$ is the set of the module's inputs.
- $O_{id\_module}$ is the set of the module's outputs.
- $IS_{id\_module}$ is the set of the module's internal signals.

Let us consider an example of a module M based on a port mapping between two instances, A and B. A is a counter (countA) that counts from 0 to MaxA, its output P is set to 1 when countA = MaxA. B is a controller, which outputs change as follows:

- G=1 and R = 0 when countA = N with N < MaxA and countB = MaxB;
- G=0 and R= 1; when countA ≠ N and countB ≤ MaxB.
- G=0 and R=0; when countA = N and countB < MaxB

The resulting HLS for each instance A and B in module M are given in Fig. 3.

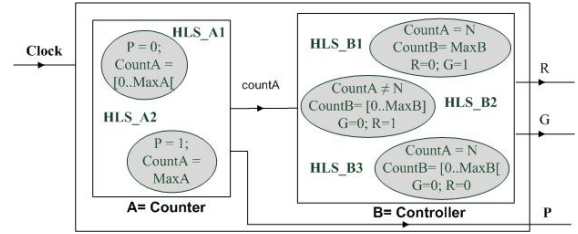

Fig. 3. HLS of two different modules, A and B. maxA, maxB and N are constants with N <MaxA

In the proposed methodology we identify even and odd numbers in order to avoid as much as possible illegal values. Here is a VHDL example:

```
Consider 4-bit signals.
 x <= inPort; // read x
 y <= 4 * x mod 10; // y never takes odd
values
 z <= y + 2; // also z
```

For this example there will be one HLS given that no assignments are done under conditions. Possible valid values of x,y and z can be combined together under all conditions. The HLS for this example will be defined as follow:
X=[0..15] Y=2*I with I=[0..5] and Z=2*J with J=[1..6].

## II. PROPOSED METHODOLOGY APPLIED ON A VHDL EXAMPLE

The following will describe intermediate and final results for every step of the constraint extraction procedure applied on the design example in Fig. 1.

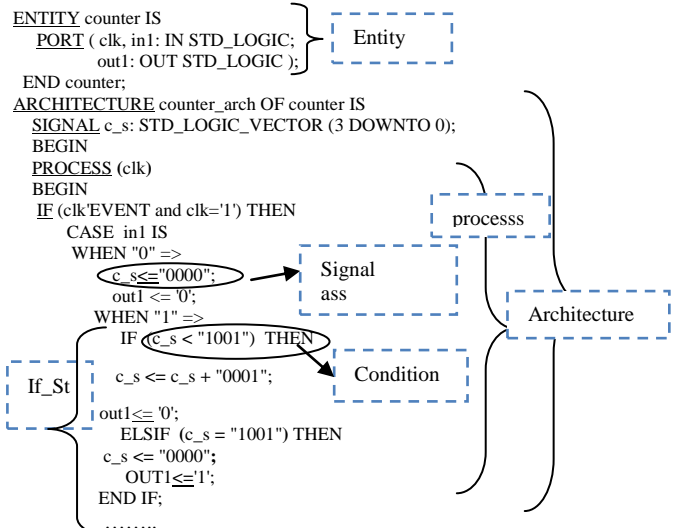

Fig. 1. A VHDL example

## 1) VHDL statement identification

Here we describe the VHDL statements of Fig.1 that are identified by the tool based on the proposed VHDL PEG. The resulting formats are shown in Fig. 2.

a)

| Entity | Id_port | Port_name | direction | Signal |
|---|---|---|---|---|
| Counter | 1 | Clk | In | n/a |
| Counter | 2 | In1 | In | n/a |
| Counter | 3 | Out1 | Out | n/a |

b)

| Component_name | Id_port | Direction |
|---|---|---|

c)

| Signal | Type |
|---|---|
| In1 | Std_logic |
| Out1 | Std_logic |
| C_s | Std_logic _vector (3 downto 0) |

d)

| Id_instance | Component_name | Id_port | Connected to |
|---|---|---|---|

e)

| Id_sa | Signal | Id_cond | Id_op | expression |
|---|---|---|---|---|
| 1 | C_s | 1, 2 | 3 | C_s<=0000 |
| 2 | Out1 | 1, 2 | 4 | Out1<=0 |
| 3 | C_s | 3, 4 | 7 | C_s<=c_s+1 |
| 4 | Out1 | 3, 4 | 8 | Out1<=0 |
| 5 | C_s | 3, 5, 6 | 11 | C_s<=0000 |
| 6 | Out1 | 3, 5, 6 | 12 | Out1<=1 |

f)

| Id_case | Id_cond | Condition |
|---|---|---|
| 1 | 2 | In1= 0 |
| 1 | 3 | In1= 1 |
| 1 | 7 | Others |

g)

| Id_if | Id_cond | Condition |
|---|---|---|
| 1 | 1 | Clk= 1 and clk'event |
| 2 | 4 | C_s<1001 |
| 2 | 5 | C_s≥1001 |
| 3 | 6 | C_s=1001 |

h)

| Id_cond | Id_op | Signal | operator |
|---|---|---|---|
| 1 | 1 | Clk | = |
| 2 | 2 | In1 | = |
| 3 | 5 | In1 | = |
| 4 | 6 | C_s | < |
| 5 | 9 | C_s | ≥ |
| 6 | 10 | C_s | = |
| 7 | n/a | n/a | null |

i)

| Id_op | cond\|Sa | Operation | Operand | Id_hlstate |
|---|---|---|---|---|
| 1 | Cond | Clk= 1 and clk'event | Clk | 1,2,3,4,5 |
| 2 | Cond | 0 | In1 | 1 |
| 3 | Sa | 0000 | C_s | 1 |
| 4 | Sa | 0 | Out1 | 1 |
| 5 | Cond | 1 | In1 | 2, 3, 4, 5 |
| 6 | Cond | 1001 | C_s | 3 |
| 7 | Sa | C_s + 1 | C_s | 3 |
| 8 | Sa | 0 | Out1 | 3 |
| 9 | Cond | 1001 | C_s | 4 |
| 10 | Cond | 1001 | C_s | 5 |
| 11 | Sa | 0000 | C_s | 5 |
| 12 | Sa | 1 | Out1 | 5 |

Fig. 2. VHDL statement representations and their dependencies based on the example in Fig.1. a) entity b) component c) signal type d) instance e) signal assignment f) case statement g) if statement h) condition i) operation

## 2) State signal initial values computation

The initial range of valid values (min and max values allowed), are computed depending on the respective signal types (Fig. 2.c), and then are stored, as shown in Table I. The set of initial legal state signal values $LV_0$ is computed as follows:

$$LV_0 = \left\{ \bigcup_{j=0}^{N} \left[ x_{j\min}, x_{j\max} \right] \right\}; \text{ with } j \in \{0..N\}.$$

Based on the example in Fig. 1:

$$LV_0 = \{[0,1], [0,15], [0,1]\};$$

TABLE I. INITIAL STATE SIGNAL VALUES CORRESPONDING TO THE EXAMPLE IN FIG. 1

| Signal identifier | Min | Max |
|---|---|---|
| In1 | 0 | 1 |
| C_s | 0 | 15 |
| Out1 | 0 | 1 |

## 3) High Abstraction Level State identification

As mentioned earlier, an HLS is the set of state signal ranges of values that can appear simultaneously in the design under the same conditions. It corresponds to the values extracted from signal assignments, under the same conditions, as well as the actual condition values. Each HLS is therefore characterized by its constraints, which consist of the set of conditions; and its effects, which consist of the signal assignments under these conditions. In this step, we identify the set of constraints and effects of each HLS.

Note that the synchronization condition (if clk'event and clk=1) does not define an HLS, and is thus not considered in the HLS identification process.

When a signal assignment is identified, the current *id_hlstate* value is assigned and stored in its corresponding representation. Each signal assignment is assigned to one specific HLS while each condition may be assigned to different HLS. To assign conditions to their corresponding HLS, we keep track of all the active conditions, and once a new HLS is identified, the list of active conditions is assigned to the new HLS, as shown in Table II. Based on this table, we deduce the corresponding HLS *id* for each condition and complete the table in Fig. 2.i.

TABLE II. HLS TABLE COMPUTED BASED ON THE EXAMPLE IN FIG. 1

| Id_state | Constraints |
|---|---|
| 1 | {1,2} |
| 2 | {1,3} |
| 3 | {1,3,4} |
| 4 | {1,3,5} |
| 5 | {1,3,5,6} |

*4) Implementation details*

The module legal HLS extraction is automated with the COMPUTE procedure. The procedure evaluates each of the modules operations and updates the state signal constraints (Table A.III) and state signal assignment (Table A.IV) tables. The result is the module set of legal HLS, $Q_{id\_module}$ and the legal HLS table (Table A.V).

The characteristics of the main procedure, *COMPUTE,* are:
- *Inputs:*
  - The module signal assignment, condition, and operation representations (Fig. 2.e, 2.h, 2.i*)*.
  - The list of initial state signal values $LV_0$ computed according to signal types (section B.1).

$$LV_0 = \left\{ \bigcup_{j=0}^{N} \left[ x_{j\min}, x_{j\max} \right] \right\}$$

In the example in Fig. 1, $LV_0 = \{[0,1],[0,15],[0,1]\}$

- Output:
  - The set of legal HLS values $Q_{id\_module}$ of the module:

  $$Q_{id\_module} = \{S_0,...,S_{id\_hlstate},...,S_S\} \text{ with}$$

  $S_{id\_hlstate} = \bigcup_{j=o}^{N}$ range of values $x_j$, and $x_j$ is a state signal

  $x_j \in \{In_{id\_module}, IS_{id\_module}, Out_{id\_module}\}$ where:
  - $In_{id\_module}$ is the set of the module inputs.
  - $O_{id\_module}$ is the set of the module outputs.
  - $IS_{id\_module}$ is the set of the module internal signals.
  - The legal HLS table modeling the set of legal HLS for each module. For each *id_hlstate* in the table, we assign a range of values corresponding to each design state signal.

TABLE III. RESULTING STATE SIGNAL CONSTRAINTS FOR THE EXAMPLE IN FIG.1

| Id_op | Id_hlstate | Cond \| SA | Signal identifier | Min | max | I\|O\|IS |
|---|---|---|---|---|---|---|
| 2 | 1 | Cond | In1 | 0 | 0 | I |
| 5 | 2,3,4,5 | Cond | In1 | 1 | 1 | I |
| 6 | 3 | Cond | C_s | 0 | 8 | IS |
| 9 | 4 | Cond | C_s | 9 | 15 | IS |
| 10 | 5 | Cond | C_s | 9 | 9 | IS |

TABLE IV. RESULTING STATE SIGNAL ASSIGNMENTS FOR THE EXAMPLE IN FIG.1

| Id_op | Id_hlstate | Cond \| SA | Signal identifier | Min | Max | I\|O\|IS |
|---|---|---|---|---|---|---|
| 3 | 1 | SA | C_s | 0 | 0 | IS |
| 4 | 1 | SA | Out1 | 0 | 0 | O |
| 7 | 3 | SA | C_s | 1 | 9 | IS |
| 8 | 3 | SA | Out1 | 0 | 0 | O |
| 11 | 5 | SA | C_s | 0 | 0 | IS |
| 12 | 5 | SA | Out1 | 1 | 1 | O |

TABLE V. RESULTING LEGAL HLS FOR THE EXAMPLE IN FIG. 1

| Id_hlstate | I | IS | O |
|---|---|---|---|
|  | In1 | C_s | Out1 |
| 1 | Min | Max | Min | Max | min | max |
| 1 | 0 | 0 | 0 | 0 | 0 | 0 |
| 2 | 1 | 1 | 1 | 9 | 0 | 0 |
| 3 | 1 | 1 | 0 | 0 | 1 | 1 |

## III. DESIGN'S HIERARCHY ANALYSIS EXAMPLE

The third step in the process is a hierarchy analysis based on the extracted information of the first step.

*Module connectivity analysis*

We consider the design in Fig. 3. The corresponding module and instance representations built based on VHDL parsing are shown in TABLES V-VIII.

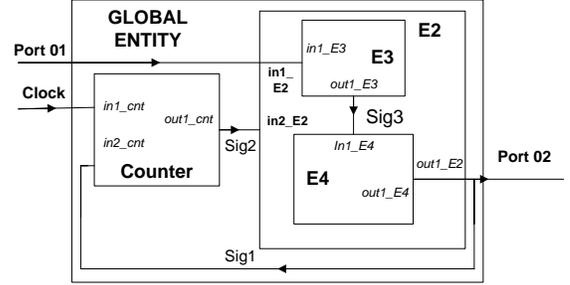

Fig. 3. Example of a design composed of several components

TABLE VI. ENTITY REPRESENTATIONS CORRESPONDING TO THE DESIGN IN FIG. 3

| Entity_name | Id_port | Port_name | Direction | Sig. |
|---|---|---|---|---|
| Global entity | 1 | Clock | In | n/a |
| Global entity | 2 | Port01 | In | n/a |
| Global entity | 3 | Port02 | Out | Sig1 |
| Counter | 1 | Clock | In | n/a |
| Counter | 2 | In1_count | In | n/a |
| Counter | 3 | Out1_count | Out | n/a |
| E2 | 1 | In1_E2 | In | n/a |
| E2 | 2 | In2_E2 | In | n/a |
| E2 | 3 | Out1_E2 | Out | n/a |
| E3 | 1 | In1_E3 | In | n/a |
| E3 | 2 | Out1_E3 | Out | n/a |
| E4 | 1 | In1_E4 | In | n/a |
| E4 | 2 | Out1_E4 | Out | n/a |

TABLE VII. GLOBAL ENTITY INSTANCE REPRESENTATIONS CORRESPONDING TO THE DESIGN IN FIG. 3

| Id_instance | Component_name | Id_port | Signal |
|---|---|---|---|
| 1 | Counter | 1 | Clock |
| 1 | Counter | 2 | Sig1 |
| 1 | Counter | 3 | Sig2 |
| 2 | E2 | 1 | Port01 |
| 2 | E2 | 2 | Sig2 |
| 2 | E2 | 3 | Sig1 |

TABLE VIII. E2 INSTANCE REPRESENTATIONS CORRESPONDING TO THE DESIGN IN FIG. 3

| Id_instance | Component_name | Id_port | Signal |
|---|---|---|---|
| 3 | E3 | 1 | In1_E2 |
| 3 | E3 | 2 | Sig3 |
| 4 | E4 | 1 | Sig3 |
| 4 | E4 | 2 | Out1_E2 |

Based on these tables, we perform a connectivity analysis. A data structure is built as shown in TABLE XIII. In this structure, each row represents a different instance, and for each instance, we define its hierarchy level, as well as all its input dependencies and its output dependencies; in cases where they are connected to the parent module output, this information will be useful when flattening the model.

TABLE IX. DATA STRUCTURE BUILT AFTER CONNECTIVITY ANALYSIS OF THE DESIGN IN FIG. 7

| Inst | Module | Level | In_dependencies | Out_dependencies |
|---|---|---|---|---|
| 1 | Counter | 1 | In1= Clock<br>In2= Out1(2) | |
| 2 | E2 | 1 | In1= port01<br>In2= Out1(1) | Out1= Port02 |
| 3 | E3 | 2 | In1= In1(2) | |
| 4 | E4 | 2 | In1=out1(3) | Out1=out1(2) |

*Flatten model*

Once we have the module's hierarchical and connectivity information, we build a flatten model to represent direct relations among instances as follow :

---

For each instance $I_i$ with $i \in \{0..I\}$

If $(in(I_i)= out(I_j))$ and $I_i$ has a submodule $I_k$ or/and $I_j$ has a submodule $I_l$ with:
- $(in(I_k)= in(I_i))$ then $in(I_k)= out(I_j)$
- $(out(I_l)= out(I_j))$ then $in(I_k)= out(I_l)$
- $(in(I_k)= in(I_i)$ and $(out(I_l)= out(I_j))$ then $in(I_k)= out(I_l)$

If $(out(I_i)= out(I_j))$ and $I_i$ has a submodule $I_k$ with
- $(out(I_k)= out(I_i))$ then $Out(I_k)= out(I_j)$

---

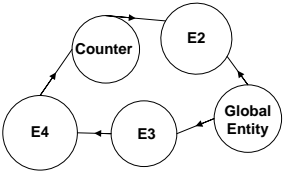

| Inst. | In_dependencies |
|---|---|
| 1 | In1= Clock<br>In2= Out1(4) |
| 2 | In1= Port01<br>In2= Out1(1) |
| 3 | In1= Port01 |
| 4 | In1=out1(3)<br>Out1= port02 |

Fig. 4. The corresponding flatten model of the design shown in the example in Fig. 3.

## IV. DESIGN LEGAL HLS EXTRACTION

To compute the final design set of legal HLS values, we need to combine the set of legal HLS of all the instantiated modules while considering instance connections and dependencies due to module connectivity.

In this methodology, all combined sets are built based on instance connections while respecting HLS dependencies and avoiding having illegal states built as follow:

```
For each HLS_i(A) ∈ Q_A
 For each HLS_j(B) ∈ Q_B
       X= Out (A) ∩ In (B)
//where out(M) is the set of possible values of
the output/input of the module M
  If X = ∅  then
  Next;
  Else
```

$$HLS(A, B) = \bigcup_{\min}^{\max} \{In(A), IS(A), X, IS(B), out(B)\}$$

```
  Q(A,B)= Q(A,B) U HLS(A,B)
    End if;
  End for;
End for
```

For instance, while computing the combined HLS of instances A (HLS_A1, HLS_A2) and B (HLS_B1, HLS_B2, HLS_B3) described in Fig. 3, the dependencies between HLS are taken into account. Fig. 5 describes the resulting HLS values set. It is composed of 4 HLS instead of 6. HLS_A2 could not be combined with HLS_B1 and HLS_B3 because of the conflicting values of countA in each of the states. Likewise, while combining HLS_A1 and HLS_B2, the range of countA is adapted to cover the legal values and prevent illegal states from being built.

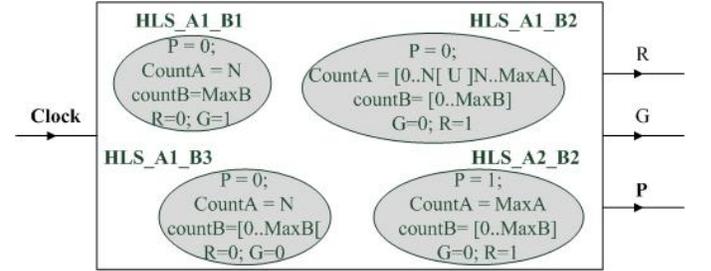

Fig. 5. Resulting HLS set of module M described in Fig. 4

## REFERENCES

[1] C. Hobeika, C. Thibeault and J.F Boland, "Functional Constraint Extraction From Register Transfer Level for ATPG," *TVLSI, submitted September 2013*.